\documentclass[journal,comsoc]{IEEEtran}
\UseRawInputEncoding
\usepackage{setspace}
\usepackage{cite}
\ifCLASSINFOpdf
	\usepackage[pdftex]{graphicx}
\else
	\usepackage[dvips]{graphicx}
\fi
\usepackage{amsmath,amssymb,amsthm}
\usepackage{caption}
\usepackage{amsfonts}
\usepackage{algorithm,algorithmic}
\usepackage{array}
\usepackage{makecell}				
\usepackage{cite}
\usepackage{esint}
\usepackage{enumerate}
\usepackage{times}
\usepackage{url}
\usepackage{stfloats}
\usepackage{multirow}
\usepackage{subfigure}
\ifCLASSOPTIONcompsoc
  \usepackage[caption=false,font=normalsize,labelfont=sf,textfont=sf]{subfig}
\else
  \usepackage[caption=false,font=footnotesize]{subfig}
\fi

\newcommand{\PreserveBackslash}[1]{\let\temp=\\#1\let\\=\temp}
\newcolumntype{C}[1]{>{\PreserveBackslash\centering}p{#1}}
\newcolumntype{R}[1]{>{\PreserveBackslash\raggedleft}p{#1}}
\newcolumntype{L}[1]{>{\PreserveBackslash\raggedright}p{#1}}

\begin{document}
%
\title{Beam Management in Ultra-dense mmWave Network via Federated Reinforcement Learning: An Intelligent and Secure Approach}

\author{Qing~Xue,~\IEEEmembership{Member,~IEEE,}
        Yi-Jing~Liu,
        Yao~Sun,~\IEEEmembership{Senior Member,~IEEE,}
        Jian~Wang,
        Li~Yan,~\IEEEmembership{Member,~IEEE,}
        Gang~Feng,~\IEEEmembership{Senior Member,~IEEE,}
        and Shaodan~Ma,~\IEEEmembership{Senior Member,~IEEE}
\thanks{Manuscript received XXX XX, 2021; revised XXX XX, 202X. Parts of this paper were presented at the 2021 IEEE Global Communications Conference (GLOBECOM) \cite{Globecom-BMFL}.}

\thanks{Q. Xue is with the School of Communication and Information Engineering, Chongqing University of Posts and Telecommunications (CQUPT), Chongqing 400065, China, and with the National Key Laboratory of Science and Technology on Communications, University of Electronic Science and Technology of China (UESTC), Chengdu 611731, China, and also with the State Key Laboratory of Internet of Things for Smart City, University of Macau, Macao SAR, China (e-mail: xueq@cqupt.edu.cn).}

\thanks{Y. Liu, J. Wang and G. Feng are with the National Key Laboratory of Science and Technology on Communications, UESTC, Chengdu 611731, China (e-mails: liuyijing@std.uestc.edu.cn, wangjians@std.uestc.edu.cn, fenggang@uestc.edu.cn).}

\thanks{Y. Sun is with James Watt School of Engineering, University of Glasgow, UK (e-mail: Yao.Sun@glasgow.ac.uk).}

\thanks{L. Yan is with the Key Lab of Information Coding \& Transmission, Southwest Jiaotong University, Chengdu 610031, China (e-mail: liyan@swjtu.edu.cn).}

\thanks{S. Ma is with the State Key Laboratory of Internet of Things for Smart City and the Department of Electrical and Computer Engineering, University of Macau, Macao SAR, China (e-mail: shaodanma@um.edu.mo).}
}

\maketitle

\begin{abstract}
Deploying ultra-dense networks that operate on millimeter wave (mmWave) band is a promising way to address the tremendous growth on mobile data traffic. However, one key challenge of ultra-dense mmWave network (UDmmN) is beam management due to the high propagation delay, limited beam coverage as well as numerous beams and users. In this paper, a novel systematic beam control scheme is presented to tackle the beam management problem which is difficult due to the non-convex objective function. We employ double deep Q-network (DDQN) under a federated learning (FL) framework to address the above optimization problem, and thereby fulfilling adaptive and intelligent beam management in UDmmN. In the proposed beam management scheme based on FL (BMFL), the non-raw-data aggregation can theoretically protect user privacy while reducing handoff cost. Moreover, we propose to adopt a data cleaning technique in the local model training for BMFL, with the aim to further strengthen the privacy protection of users while improving the learning convergence speed. Simulation results demonstrate the performance gain of our proposed scheme.
\end{abstract}


\IEEEpeerreviewmaketitle

\section{Introduction}
The past few years have witnessed the explosive growth of wireless data traffic, and this growth trend will continue \cite{Cisco} due to the fast development of mobile multimedia applications and Internet of Things. With beamforming and massive multiple-input multiple-output (MIMO) \cite{Two-Way-Massive-MIMO-Relaying,Tensor-mmWave-MIMO,Matrix-Monotonic-Optimization}, millimeter wave (mmWave) communication has been widely acknowledged as a promising means to meet the projected requirements by dint of the abundant spectrum resources. However, compared with the traditional microwave communication networks, mmWave communication networks face two critical challenges. One is the limited coverage because of the serious propagation path loss. Ultra-dense network (UDN) \cite{ultra-dense-networks} is an effective technique to address this issue, where various small cell base stations (SBSs) with different coverage are densely deployed, and thus the distance between users and SBSs becomes closer. The other is the susceptible to blockage due to the inherent directivity. An efficient strategy for enabling reliable transmissions and enhanced data rates is to employ the multi-connectivity technique \cite{Multi-Connectivity}, which enables a user to connect to several SBSs simultaneously. In recent years, intelligent reflecting surface (IRS) \cite{Large-Intelligent-Surface-Antennas-LISA,Unified-IRS-Aided-MIMO,Rate-Analysis-RIS-Assisted-MIMO,Sum-Rate-RIS-Assisted-MIMO} has also become a powerful means to overcome this plight. Meanwhile, heterogeneous networks (HetNets) have been treated as one promising candidate for universal coverage and enhancing overall capacity, where SBSs are deployed underlying a core macro network. Inspired by these research results, we consider ultra-dense mmWave HetNets with multi-connectivity in this study, and focus on the issue of beam management since it affects the system performance principally.

Due to hardware constraints, each mmWave SBS (mSBS) may only support limited beams simultaneously. However, for an ultra-dense mmWave system, the number of operating beams could be extremely large caused by the densely deployed mSBSs. This leads to a more complex and critical beam management problem than that in conventional non-dense mmWave systems. In current mmWave systems, the term beam management is usually known as fine alignment of the transmitter and receiver beams to perform a variety of control tasks including initial access for idle users and beam tracking for connected users \cite{Tutorial-Beam-Management}. In this work, in order to improve beam utilization and reduce inter-beam interference, we manage beams in a systematic manner instead of beam-by-beam basis. The systematic beam management mainly refers to the dynamic control of beam directions at the mSBS side (\emph{i.e.}, mSBS beam configuration) based on periodically sensing instantaneous user distributions.

Moreover, we introduce machine learning (ML) technique to realize intelligent and proactive beam management. Although the existing ML-based mechanisms bring a number of benefits in beam management, they also expose some potential risks, most typically, in security and privacy protection. It is because that traditional centralized ML techniques normally require data collection and processing by a central controller, but the training data may be privacy sensitive in nature. This problem is becoming a bottleneck of large-scale implementation of traditional centralized ML schemes in practical applications. In addition, the overhead caused by centralized data aggregation and processing is usually significant. These reasons have led to a growing interest in a new ML model, namely federated learning (FL) \cite{Federated-Learning}. In FL, participating learners collaboratively train a shared model by exploiting their local computation capability and data, and thus only local model updates instead of raw data need to be transferred to a centralized model aggregation server. Thereby, FL can be exploited to train ML models in a distributed way while preserving user privacy. These natures of FL motivate this work, which is the first time to exploit FL for designing mmWave beam management scheme in the open literature. Unfortunately, there are numerous mobile users in ultra-dense mmWave networks and the computing resources occupied for training on a specific mSBS may be inadequate. Hence, directly applying legacy FL framework to solve the  the systematic beam management problem may not always effective and  obtain satisfactory performance. In this paper, before model training, mSBSs will clean the training data according to actual demand, so as to ensure the privacy of specific users while improving the training efficiency.

\subsection{Prior Work}
Some recent papers \cite{Tutorial-Beam-Management,Modular-High-Resolution,Standalone-Non-Standalone,Beam-Management-5G-Beyond} have provided overview of beam management for mmWave in 5G New Radio standard. Beam management procedures for handling mobility can be categorized into beam sweeping, beam measurement and reporting, beam determination, beam maintenance, and beam failure recovery \cite{Beam-Management-5G-Beyond}. To date, most of the investigations in beam management tackle the problem by resorting to beam training, sparse channel estimation, and location aided beamforming \cite{Beam-Management-Q-Learning}, while beam training is most commonly used. In beam training, both ends of a link search through the set of available beams in either an exhaustive or iterative mode until a good link is established. In particular, a fixed sector-level beam training is specified in IEEE 802.11ad/ay for initial access \cite{802.11ay}. A \emph{sector} is defined as a specific broad antenna radiation pattern. For mobile users, the beam training procedure should be performed frequently to ensure the accurate acquisition of the channel state information (CSI). But there exist a number of problems for this process, including high complexity, significant training overhead, and access delays. Motivated by this, the authors in \cite{Beamspace-SU-MIMO} and \cite{Beam-Management-Beamspace-MU-MIMO} investigated the challenges and potential solutions of
downlink beamspace SU-/MU-MIMO including multi-beam training, cooperative beam tracking, and multi-beam power allocation. To solve the problem of co-channel interference in dense mSBS scenario, a large-scale CSI-based interference coordination approach is proposed in \cite{Scalable-Interference-Coordination}. To improve the training efficiency especially in dynamic environments, the priori-aided beam training termed beam/channel tracking in the literatures is of crucial importance for users to maintain seamless connectivity \cite{Tracking-angles-departure-arrival,Robust-Beam-Tracking,Codebook-Based-Training}. Nevertheless, beam training/tracking is generally used to ensure a specific user's quality of service (QoS) requirements. Moreover, for most practical scenarios, the assumption of mmWave channel modeling required by channel tracking techniques is too stringent to meet \cite{beam-Bandit-Learning}.

As mobile environments are increasingly complex, heterogeneous, and evolving, ML techniques have attracted significant attention to optimize wireless communication systems in the last few years, owning to its ability in creating smart systems that can take sequential decisions and make accurate predictions. Prior work has established that ML is a good tool for beam management in mmWave communication systems, especially in mobile applications or dynamic environments \cite{sun2017smart,beam-Bandit-Learning,Deep-Learning-Coordinated-Beamforming,Deep-Learning--Beam-Management-Interference,Deep-learning-channel-estimation-tracking,User-centric-Association}. For example, a deep learning based coordinated beamforming algorithm is proposed in \cite{Deep-Learning-Coordinated-Beamforming} to reduce the training overhead. In \cite{Deep-Learning--Beam-Management-Interference}, the authors proposed a deep neural network (DNN)-based beam management and interference coordination algorithm to reduce the interference and improve the sum-rate of dense mmWave network. In our previous work \cite{User-centric-Association}, a deep Q-network based user-centric association scheme is designed to provide reliable connectivity and high achievable data rate for ultra-dense mmWave networks. Furthermore, some work as in \cite{FL-Edge-Computing} and \cite{RAN-Slicing-FL} uses FL to address their corresponding research issues. Intuitively, FL framework can further enhance the performance of the adaptive beam management scheme in terms of security and privacy protection, but the relevant topic is still in the exploratory research stage.

\subsection{Contributions}
In this work, we first employ federated reinforcement learning to realize intelligent and proactive beam configuration (on the mSBS side), named BMFL, and then adopt a $\mu$Wave-assisted multiple association \cite{User-centric-Association} to ensure user QoS in the ultra-dense mmWave network (UDmmN). The key contributions can be summarized as follows.

\hangafter 1
\hangindent 1.0em
\noindent
$\bullet$ Different from the conventional beam management schemes based on the optimal pairing of transmit and receive beams, i.e., beam training and/or tracking, we present a systematic beam management scheme that pays more attention to the global performance of UDmmN.

\hangafter 1
\hangindent 1.0em
\noindent
$\bullet$ Considering the limited ability of mSBS to support simultaneous beams, we propose to perform dynamic beam configuration by periodically sensing instantaneous user distributions to improve beam utilization, rather than static beam deployment as in traditional scenarios.

\hangafter 1
\hangindent 1.0em
\noindent
$\bullet$ We address the issue of data privacy in BMFL by using an FL framework to avoid any exchange of user private information (such as location, trajectory, behavior). Different from existing work, applying federated DRL to beam management of UDmmN is the first attempt, to the best of the authors' knowledge.

\hangafter 1
\hangindent 1.0em
\noindent
$\bullet$ According to the coverage of mSBS and the frequency of user participation in training, we propose to employ data cleaning technique in the BMFL algorithm to further strengthen the privacy protection of specific users while improving the learning convergence speed.

The remainder of this paper is organized as follows. System model of the UDmmN is described in Section II. Systematic beam management problem is formulated in Section III. In Section IV, algorithm of BMFL is presented. Performance of BMFL is evaluated in Section V. Finally, Section VI concludes the paper.

\begin{figure}[htbp]
  \begin{center}
    \scalebox{0.54}[0.54]{\includegraphics{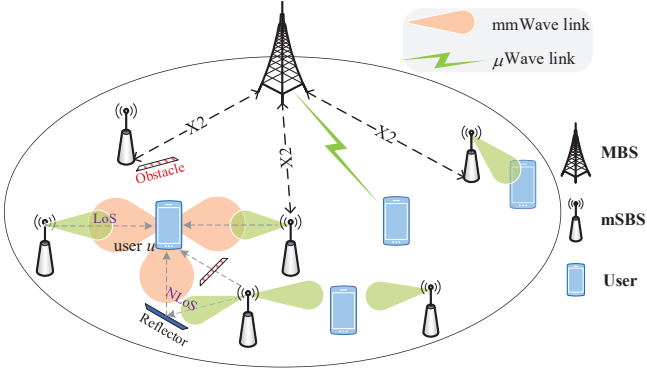}}
    \caption{An illustration of ultra-dense mmWave network.}
    \label{fig:1}
  \end{center}
\end{figure}

\section{System Model}
Fig.~\ref{fig:1} illustrates a two-tier heterogeneous UDmmN, in which ultra-dense mSBSs are deployed randomly under the coverage of one macro BS (MBS) operating on conventional microwave band. In order to communicate and exchange control information, the MBS and mSBSs are inter-connected via traditional backhaul X2 interfaces. Meanwhile, multi-connectivity, which is a feature that each user maintains multiple possible signal links to different cells, is introduced to improve the robustness of mmWave communications. Due to the short distance and high directivity of mmWave transmission, a full reuse of mmWave spectrum usually does not suffer from serious interference. Hence, we assume that all mSBSs share the total mmWave bandwidth $W_{mm}$. All mSBSs are denoted by ${\cal B} = \left\{ {1,...,B} \right\}$ and the users moving randomly within the UDmmN are denoted by ${\cal U} = \left\{ {1,...,U} \right\}$, where $B=\left| {\cal B} \right|$ and $U=\left| {\cal U} \right|$.

\begin{figure}[htbp]
  \begin{center}
    \scalebox{0.65}[0.65]{\includegraphics{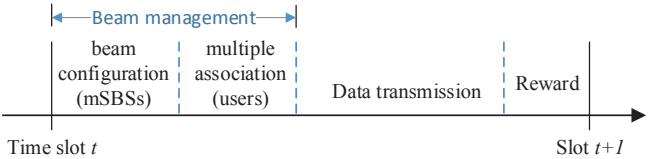}}
    \caption{Beam management operations in time slot $t$.}
    \label{fig:2}
  \end{center}
\end{figure}

Due to the hardware limitation, we assume that mSBS $b$ ($b \in {\cal B}$) can form up to $M_b$ transmit beams simultaneously by adopting beamforming technique. In order to compensate the high propagation loss of mmWave signals, the width of each beam is generally narrow. Consequently, the small cell may not be fully covered by these beams. For ease of illustration, we divide the small cell $b$ ($b \in {\cal B}$) into $S_b$ transmit sectors (or beam directions) with respect to the condition $0 < {M_b} \le {S_b}$. We assume that the beams covering different sectors are mutually orthogonal in space, and each beam can serve multiple users within its coverage, for example, in a time division multiplexing manner. Different from traditional beam management, which is used to maximize the quality of a single link through beam training and tracking, the beam management in this paper aims to maximize the system level performance through the beam configuration on the mSBS side. In order to improve beam utilization, beam management leveraging double deep Q-network (DDQN) \cite{DDQN} under an FL framework is used in our work. Meanwhile, as shown in Fig.~\ref{fig:2}, we assume that beam management is performed in a synchronous time-slotted fashion \cite{A-Learning-Approach} and the beams are static during each time slot. A time slot here is defined as a beam adjustment interval, the duration of which is generally related to user mobility and sector range. Within each slot, there are three main operations. (i) At the beginning of a slot, mSBSs calculate the accumulated network performance over previous slots, and then manage beams by adopting FL-based algorithm. (ii) Users choose suitable mSBSs to associate with. (iii) Transmit data at the selected beam/mSBS.

In UDmmN, in addition to the MBS, user $u$ ($u \in {\cal U}$) may receive data from several mSBSs surrounding it and can associate with up to $B_u^{\max }$ of them. Let ${x_{u,b}(t)} = \left\{ {0,1} \right\}$ denote the binary association indicator variable for user $u$ ($u \in {\cal U}$) and mSBS $b$ ($b \in {\cal B}$), where ${x_{u,b}(t)} = 1$ if user $u$ is associated with mSBS $b$ at time $t$, otherwise ${x_{u,b}(t)} = 0$. Denoting by ${\cal B}_u(t)$ (${{\cal B}_u(t)} \subseteq {\cal B}$) the set of mSBSs associating with user $u$ at time $t$, the number of these mSBSs is ${B_u(t)} = \left| {{{\cal B}_u}(t)} \right| = \sum\limits_{b \in {\cal B}} {{x_{u,b}(t)}}  \le B_u^{\max }$. There are two types of user association in the UDmmN. One is single-mode association, where a given user $u$ associates with the MBS only. This case will occur when ${{\cal B}_u(t)} = \emptyset$. The other is dual-mode association, where user $u$ associates with both the MBS and the mSBSs in its vicinity. For the UDmmN, we focus on mmWave association instead of the association with the MBS, as the user association problem in conventional microwave communication networks has been studied extensively. We assume that users in the UDmmN are always associated with the MBS to ensure seamless communication. To ensure the QoS requirements of users, the mmWave association is carried out after the beam configuration.

Moreover, the propagation models in this study are as follows. For microwave transmission, the channel gain is \cite{Microwave-channel-model}
\begin{equation}
h\left( {{d_{mbs}}} \right) = \kappa  + 10\rho {\log _{10}}\left( {{d_{mbs}}} \right),
\end{equation}
where $d_{mbs}$ is the distance from the user to the MBS, $\kappa$ is the path loss factor (in dB) per meter, and $\rho$ is the path loss exponent. When user $u$ is associated with the MBS, the signal-to-interference-plus-noise ratio (SINR) can be expressed as
\begin{equation}
{{SINR}_{u,mbs}} = \frac{{{p_u^0} \cdot h\left( {{d_{mbs}}} \right)}}{{{p_N^0} + {I_u}}},
\end{equation}
where $p_u^0$ is the transmission power of MBS allocated to user $u$, $p_N^0$ denotes the receiver noise power, and $I_u$ denotes the cochannel interference. For mmWave transmission, the path loss can be modeled as \cite{Channel-Modeling}
\begin{equation}
{PL}\left( d_{u,b} \right) = \alpha  + 10\beta {\log _{10}}\left( d_{u,b} \right) + \xi \left[ {{\rm{dB}}} \right],\xi  \sim N\left( {0,{\sigma ^2}} \right),
\end{equation}
where $d_{u,b}$ is the distance between user $u$ and mSBS $b$ in meters, $\alpha$ and $\beta$ are the least square fits of floating intercept and slope over the measured distances (30 to 200 m), and $\sigma ^2$ is the lognormal shadowing variance. The values of $\alpha$, $\beta$ and $\sigma ^2$ are different for Line-of-Sight (LoS) and Non-LoS (NLoS) states. In general, the received power at user $u$ when associated with mSBS $b$ can be written as $p_{u,b}^R = \frac{{{p_{u,b}} \cdot G_{u,b}^T \cdot G_{u,b}^R}}{{PL\left( {{d_{u,b}}} \right) \cdot {{CF}_{u,b}}}}$, where $p_{u,b}$ is the allocated transmit power of mSBS $b$ to user $u$, $G_{u,b}^T$ and $G_{u,b}^R$ are the transmit and receive antenna gain respectively, and ${CF}_{u,b}$ is the small-scale channel fading \cite{Scalable-Interference-Coordination}. Considering that this work focuses on the beam management at the mSBS side to maximize user coverage, the small-scale channel fading is assumed to have little impact on the network-level beam adjustment. Hence, the observed SINR can be expressed as
\begin{equation}
SIN{R_{u,b}} = \frac{{{{{p_{u,b}} \cdot G_{u,b}^T \cdot G_{u,b}^R} \mathord{\left/
 {\vphantom {{{p_{u,b}} \cdot g_{u,b}^T \cdot g_{u,b}^R} {PL\left( {{d_{u,b}}} \right)}}} \right.
 \kern-\nulldelimiterspace} {PL\left( {{d_{u,b}}} \right)}}}}{{{p_N} + \sum\limits_{k \in {\cal B},k \ne b} {\frac{{{p_{u,k}} \cdot G_{u,k}^T \cdot G_{u,k}^R}}{{PL\left( {{d_{u,k}}} \right)}}} }},
\end{equation}
where $p_N$ is the noise power, and the right part of the denominator represents the total power of interfering signals.

A summary of key notations is presented in TABLE I.

\begin{table}[htbp]
\centering
\caption{Summary of key notations.}
\scalebox{0.9}{
\begin{tabular}{l|l}
\Xhline{0.5pt}
\Xhline{0.5pt}
\textbf{Notation} & \textbf{Description}\\
\Xhline{0.5pt}
\Xhline{0.5pt}

${\cal B}$     & Set of total mSBSs in the UDmmN\\
${\cal U}$    & Set of total users in the UDmmN\\
$W_{mm}$  & Total mmWave bandwidth\\
$W_{mbs}$    &  Total available bandwidth of the MBS\\
${\cal B}_u (t)$  & Set of the serving mSBSs of user $u$ at time $t$\\
$x_{u,b}(t)$     & Binary association indicator variable \\
$M_b$     & Number of beams of mSBS $b$\\
$S_b$     & Number of transmit sectors of mSBS $b$\\
$B_u^{\max}$  & Maximum number of mSBSs associated with user $u$\\
$B_u(t)$  & Number of the serving mSBSs of user $u$ at time $t$\\
$N_b(t)$  & Number of users served by mSBS $b$ at time $t$\\
$d$       & Distance between MBS/mSBS and user\\
$h(d)$            & Microwave channel gain\\
${PL}(d)$            & MmWave path loss model\\
$G\left( {\varphi _{u,b}^T} \right)$  & Transmit antenna gain between user $u$ and mSBS $b$\\
$G\left( {\varphi _{u,b}^R} \right)$  & Receive antenna gain between user $u$ and mSBS $b$\\
$p_u^0$  & Transmit power of the MBS to user $u$\\
$p_{u,b}$  & Transmit power of mSBS $b$ to user $u$\\
$\chi$   & SINR threshold\\

\Xhline{0.5pt}
\Xhline{0.5pt}
\end{tabular}
}
\end{table}

\section{Problem Formulation and Analysis}
In this section, we formulate the problem of beam management as a long-term optimization, and then discuss the tractability of this optimization problem. For the case that the user is associated with multiple mSBSs, the achievable rate of user $u$ at time $t$ should be the sum of data rate received from all the associated mSBSs. Thus, the data rate can be given as
\begin{equation}\label{eq:rate}
\begin{split}
&r_u(t) =\\
&\begin{cases}
\sum_{b\in {\cal B}_u (t)}W_{u,b}log_2\left(1+\mathit{SINR}_{u,b}(t)\right),&\text{if}~{\cal B}_u (t)\neq \emptyset,\\
\frac{W_{mbs}} {N_{mbs}(t)}log_2\left(1+\mathit{SINR}_{u,mbs}(t)\right),&\text{if}~{\cal B}_u (t)= \emptyset,
\end{cases}
\end{split}
\end{equation}
where $W_{u,b}$ is the allocated mmWave bandwith of mSBS $b$ to user $u$, $W_{mbs}$ is the total available bandwidth of the MBS, $N_{mbs}(t)$ is the number of users served by the MBS at time $t$, and $\mathit{SINR}_{u,b}(t)$ ($\mathit{SINR}_{u,mbs}(t)$) is the obtained SINR of user $u$ from mSBS $b$ (the MBS) at time $t$. Hence, the system throughput at time $t$ is
\begin{equation}
R(t) = \sum_{u\in\mathcal{U}} r_u(t).
\end{equation}

In order to improve beam utilization and reduce inter beam interference, we optimize the beam configuration of mSBSs, \textit{i.e.}, determine which sectors should be covered at time $t$ based on periodically sensing instantaneous user distributions. Denote the optimization variable $\pi_b(t)$ as the set of sectors covered by mSBS $b$ at time $t$, and the beam management policy for the whole system at time $t$ is denoted by  $\pi(t) = \left[\pi_1(t), \pi_2(t), ..., \pi_{|\mathcal{B}|}(t)\right]$. Taking a suitable policy can let more sectors be covered by mSBSs, and thus improve the
system throughput. To this end, we formulate the beam management problem as follows with the objective of maximizing the long-term system throughput.
\begin{align}
		\mathbf{P1:}~\max_{\pi(t)} &~ \lim\limits_{T \to \infty}\mathbb{E}\left[\frac{1}{T}\sum_{t} R\left( t \right)\right] \label{equ:P1} \\
	{\rm s.t.}~	
    &0 \leq |\pi_b (t)| \leq M_b, \forall b\in \mathcal{B}, t\in \mathcal{T},\tag{\ref{equ:P1}-1}\\
   &\mathit{SINR}_{u,b}(t)\geq \chi, \notag\\
   &\forall (u,b) \in \left\{(u,b)| {x_{u,b}(t)}=1\right\}, t\in \mathcal{T}, \tag{\ref{equ:P1}-2}\\
   &0 \leq \sum\limits_{b \in {\cal B}} {{x_{u,b}(t)}}  \le B_u^{\max}, \forall u\in\mathcal{U}, t\in \mathcal{T}, \tag{\ref{equ:P1}-3}\\
   & {x_{u,b}(t)} = \left\{ {0,1} \right\}, \forall u\in\mathcal{U}, b\in\mathcal{B}, t\in \mathcal{T} \tag{\ref{equ:P1}-4}
\end{align}
where $\mathbb{E}[\cdot]$ is the expectation of the variable, $\mathcal{T}$ with cardinality $T$ is the set of time slot for adjusting beam management policy, $\chi$ is the SINR threshold that users can correctly receive and decode the information, and $\hat{r}_{u}$ is the minimum requirement on data rate of user $u$. In problem \textbf{P1}, Constraint (\ref{equ:P1}-1) ensures that the maximum number of beams for mSBS $b$ is $M_b$. Constraint (\ref{equ:P1}-2) guarantees that the SINR of the link between users and the serving mSBSs should be greater than the threshold $\chi$. (\ref{equ:P1}-3) and (\ref{equ:P1}-4) are the constraints on user association, where the number of associated mSBSs for user $u$ cannot exceed the access capability $B_u^{\max}$.

Examining problem \textbf{P1} we realize that the problem is hard to solve by using traditional optimization method. The rational behind is that the long-term optimization objective with unknown user movement behavior is formulated. Thus, the network environment (including user locations, channel quality, network resources, etc.) of future time slot cannot be obtained or even mathematically modeled at the beginning. An efficient and promising way to solve \textbf{P1} is to resort to ML algorithms. Number of reinforcement learning algorithms can be adopted to solve the problem with long-term objective by interacting with dynamic environment via information exchanges. However, as mentioned above that the raw data in terms of user locations is quite private and should be carefully protected rather than being exchanged among multiple mSBSs like that in most reinforcement learning algorithms. To this effect, FL, which requires the exchanges of learning model rather than raw data, is next adopted to derive the optimal beam management policy of \textbf{P1}.

\section{FL-based Beam Management in UDmmN}
In this section, we propose a novel beam management mechanism for mSBS beam configuration based on FL in UDmmN, called BMFL, with the aim to maximize the long-term throughput while enforcing the protection of user location privacy. Specifically, we first formulate the beam management problem as a markov decision process (MDP) model, and then propose BMFL based on the MDP model by exploiting federated DRL.

In this paper, we focus on beam management in the scenario of ultra-dense mSBSs. If all the information of these mSBSs is sent to the MBS for centralized data aggregation and processing, it usually faces great challenges in terms of computation energy consumption, computational latency and learning time. Therefore, we adopt a decentralized learning technique (i.e., FL) where training datasets are distributed over mSBSs, instead of centralizing all the data. Another notable advantage of FL is the protection of data privacy. Although each mSBS can obtain the user information like location of its own serving users, it may be unconventional to share the information with other mSBSs. It is because that the location information is private to the user, it should not be shared with those mSBSs that are not associated with the user. Hence, the user information is regarded as sensitive information in the system and can be protected under the FL architecture.

Considering the problem has a high state and action dimensions, we exploit DDQN to use a neural network to estimate the value function, which improves the learning accuracy with a small compromise on the learning convergence speed. This is the insight in the DRL algorithm. The more insights lie on the FL framework, which can reduce the learning problem scale since each agent performs a distributed learning framework, and the aggregated global learning model is used to guide each agent converging fast. Also, the privacy of raw data can be ensured in the FL framework.

\subsection{Markov Decision Process Model for UDmmN}
An MDP process is composed of four-tuple ${\cal M}=({\cal S}, {\cal A}, {\cal P}, {\cal R})$, where ${\cal S}$ and ${\cal A}$ represent state and action space respectively, ${\cal P}$ is the transition probability from current state ${\cal S}_t$ to the next state ${\cal S}_{t+1}$, and ${\cal R}$ represents reward function. In our problem, a specific mSBS $b$ ($b \in {\cal B}$) makes a decision (action) on beam directions at each time slot to maximize long-term throughput and the network state may be changed by these sequential actions. We define the state, action, transition probability and reward as follows.

\hangafter 1
\hangindent 1.0em
\noindent
$\bullet$ {\emph{State}:} Current operating beam sectors and serving users of mSBSs are used to describe the system state. Specifically, ${\cal S}_t$ is the set of all network states for mSBSs at time $t$. For a specific mSBS $b$, the state is $s_t^b=\{{\cal U}_b, \pi_b(t), {\cal S}_t^{\cal K}\} \in {{\cal S}_t}$, where
${\cal U}_b$ represents the set of serving users and $\pi_b(t)$ represents the corresponding beam sectors occupied by these users. Moreover, ${\cal S}_t^{\cal K} = {\left\{ {{\pi _k}\left( t \right)} \right\}_{k \in {\cal B},k \ne b}}$ represents the available sectors of all mSBSs except for mSBS $b$.

\hangafter 1
\hangindent 1.0em
\noindent
$\bullet$ {\emph{Action}:} Let ${\cal A}_t = \{a_t^b\}$ be the set of actions for all mSBSs at time $t$. Note that an mSBS is an agent which trains local model independently. For a specific mSBS $b$, let $a_t^b=\pi_b(t)$ be the action, which means that mSBS $b$ serves users in ${\cal U}_b$ with covered beams in $\pi_b(t)$ at time $t$.

\hangafter 1
\hangindent 1.0em
\noindent
$\bullet$ {\emph{State transition probability}:}  Let the transition probability of mSBS $b$ be $P_b=Pr(s_{t+1}^b|s_{t}^b)$, which represents the probability that network state of mSBS $b$ transits from $s_t^b$ to $s_{t+1}^b$.

\hangafter 1
\hangindent 1.0em
\noindent
$\bullet$ {\emph{Reward}:} In order to maximize the long-term system throughput, we define the reward as ${\cal R}_t=R\left( t \right)$, where $R\left( t \right)$ is the optimization objective of \textbf{P1}.

In the MDP for beam management, the state for an mSBS consists of three elements, i.e., the set of serving users, the occupied beam sectors, and the available sectors of all other mSBSs. Therefore, the state space dimensions should be the combination of the number of users and the number of sectors of all mSBSs, i.e., $\left| {{{\cal U}_b}} \right| \cdot \prod\limits_{b \in {\cal B}} {{M_b}}$. Similarly, for a local agent (i.e., an mSBS), the action space dimensions can be given by $M_b$. Please note that we exploit the distributed learning scheme, so the action space is calculated for each mSBS separately.

\begin{figure}[t]
  \begin{center}
    \scalebox{0.5}[0.5]{\includegraphics{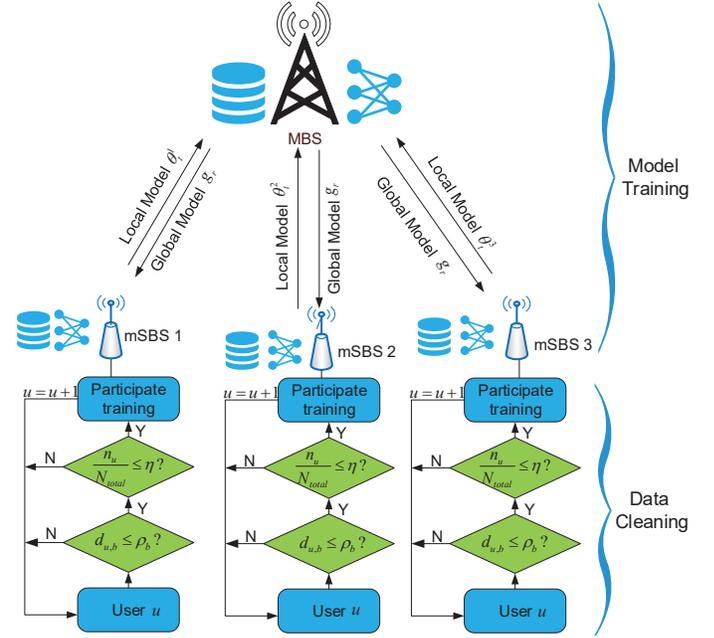}}
    \caption{Beam management based on FL (BMFL) in UDmmN.}
    \label{fig:3}
  \end{center}
\end{figure}

\subsection{FL-based Beam Management in UDmmN}
In this subsection, we propose the BMFL in UDmmN. As shown in Fig.~\ref{fig:3}, BMFL consists of two steps, {\emph{i.e.}}, data cleaning and model training (including local model updating, local model training, global model aggregating, as in \cite{RAN-Slicing-FL}). Specifically, to reduce the computing resources occupied for training, an mSBS will first clean data, \emph{i.e.}, choose training users (participants) according to the frequency of participating training and the distance between users and this mSBS. Then, to enhance user location protection while coping with large state-action space issues, DRL under an FL framework is introduced into model training. The data cleaning and model training are as follows.

\begin{figure}[t]
  \begin{center}
    \scalebox{0.5}[0.5]{\includegraphics{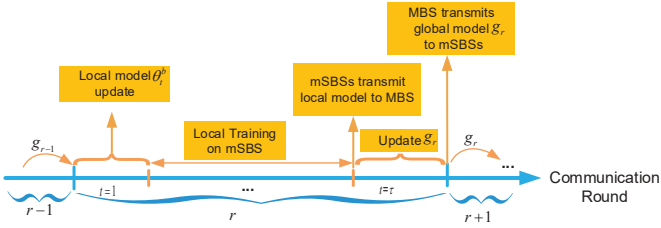}}
    \caption{The process of a communication round.}
    \label{fig:4}
  \end{center}
\end{figure}

\textbf{Data Cleaning}:
In order to ensure the quality and diversity of training data, the mSBS should perform data cleaning at the beginning of each communication round (i.e., each model update iteration). Fig.~\ref{fig:4} shows the process of a communication round, which includes five steps: local parameter initializing, local model training, local model transmission, global model updating, and global model transmission. At the begin of each communication round, mSBS will first choose the users for participating in local training.

Even if mSBSs can obtain all location information of the serving users, it is unrealistic for the mSBSs to choose all the users in their coverage range to participate in local training as 1) the computing resources occupied for training on a specific mSBS may be inadequate, 2) the location of some certain users needs to be protected. Therefore, to solve the issue mentioned-above while increasing sample diversity as much as possible, mSBS will clean data according to the following two parameters. 1) The coverage of mSBS. The users that are not located in the coverage range will not be chosen to participate in local training. 2) The frequency of participating training. If some users have not participated in local model training for a long time, mSBS will choose them as the participants in the training for next global model updating and thus to increasing the sample diversity. Therefore, the users will be chosen to train local model on the mSBS once both condition (\ref{eq:distance}) and (\ref{eq:training}) are met.
\begin{equation}\label{eq:distance}
d_{u,b}\leq \rho_b,
\end{equation}
\begin{equation}\label{eq:training}
\frac{n_u}{N_{total}}\leq \eta,
\end{equation}
where $\rho_b$ and $\eta$ are the coverage radius threshold of mSBS $b$ and the frequency threshold of participating training respectively. For a specific user $u$, $\frac{n_u}{N_{total}}$ represents the frequency of participating training, where $n_u$ is the number of participating training of user $u$ and $N_{total}$ is the total training times of the relevant mSBS so far. Here we assume ${\cal U}_b^*$ (${\cal U}_b^* \subseteq {\cal U}_b$) represents the set of the users that participate in local model training on mSBS $b$.

\textbf{Model Training}: Once finishing the data cleaning, mSBS begins to train local model including local model updating, local model training, and global model aggregating, which are shown as follows.

\emph{1) Local model updating}: We assume that each communication round consists of $\tau$ time slots. We denote the local beam management model on mSBS $b$ at time $t$ and the global model at communication round $r$ by $\theta_t^b$ and $g_r$ respectively. During each time slot, each mSBS performs local training once. At the begin of communication round $r$, mSBSs will receive global model $g_{r}$ from the MBS to update $\theta_t^b$ according to
\begin{equation}\label{eq:local}
\theta_{t+1}^{b}=g_{r}-\frac{\lambda}{K_b}\sum_{b=1}^{\left| {{\cal B}} \right|}\nabla L(\theta_{t}^{b}), 1\leq t\leq T,
\end{equation}
where $\lambda$ is the step size, $K_b$ is the total amount of training data of mSBS $b$, and $L(\theta_{t}^{b})$ is the loss function which will be given in the next part.

\emph{2) Local model training}: For a specific mSBS, once all training data is cleaned and the local model is updated, the
mSBS begins to train local beam management model based on the location information of participants within its coverage range. As mentioned, a large number of mSBSs and users result in large state space and action space. Therefore, during each communication round, we employ the discrete-action DRL algorithm, DDQN, to train the local beam management model on individual mSBSs. DDQN can tackle the issue of large state/action space by introducing the experience pool and decoupling the selection from the evaluation to reduce the correlation among data. DDQN evaluates the greedy policy according to the Q-network with weight $\theta$ and estimates state-action value $Q$ according to the target network $\hat{Q}$ with weight $\hat{\theta}$. The update in DDQN is the same as that in deep Q-network, but the target is replaced by
\begin{equation}
y_{t}^{b}={\cal R}_{t+1}+\gamma Q(s_{t+1}^{b}, \arg\max_{a_{t}^{b}}Q(s_{t+1}^{b},a_{t}^{b};\theta_t^{b});\hat{\theta_t^{b}}),
\end{equation}
where
\begin{equation}
\pi=\arg\max_{a_{t}^{b}}Q(s_{t+1}^{b},a_{t}^{b};\theta_t^{b}),
\end{equation}
is an $\epsilon$-greedy policy used to manage beam sectors, $\theta_t^{b}$ and $\hat{\theta}_t^b$ are the weight vectors of $Q$-network and $\hat{Q}$-network for mSBS $b$ respectively, and $\gamma\in \left[0,1\right]$ is the discount factor representing the discounted impact of future reward. For a specific mSBS $b$, if it is in state $s_{t}^{b}$ with action $a_{t}^{b}$ at time slot $t$, we will get the corresponding state-action value, which is given by
\begin{equation}
Q({s}_{t}^{b},{a}_{t}^{b})=\mathbb{E}[\sum_{k=t}^T \gamma^k {\cal R}_t|{s}_{t}^{b},{a}_{t}^{b}].
\end{equation}
The objective of DDQN is to minimize the gap between $Q$ and $\hat{Q}$, \emph{i.e.}, loss function. Therefore, DDQN running on each mSBS can be trained by minimizing the loss function, which is given by
\begin{equation}\label{eq:L}
L(\theta_t^{b})=\mathbb{E}[(y_{t}^{b}-Q(s_{t}^{b},a_{t}^{b};\theta_t^{b}))^{2}].
\end{equation}

Moreover, when DDQN is used to approximate the value function using the neural network, gradient descent method is employed to update the parameter value $\theta_t^{b}$. Therefore, the update scheme in DDQN is given by
\begin{equation}\label{eq:Parameter}
\begin{array}{c}{\theta_{t+1}^{b}=\theta_{t}^{b}+\alpha\left[y_t^{b}-\right.}{\left.Q\left(s_{t}^{b}, a_{t}^{b} ; \theta_t^{b}\right)\right] \nabla Q\left(s_{t}^{b}, a_{t}^{b}; \theta_t^{b}\right)}\end{array},
\end{equation}
where $\alpha$ is a scalar step size.

After training local data for $\tau$ time slots, mSBSs will send training parameters $\theta_t^{b}$ ($b \in {\cal B}$) to the MBS to update the global model.

\emph{3) Global model aggregating}: Once receiving all local models (\emph{i.e.}, $\theta_t^{b}$ for $\forall b \in {\cal B}$) at the end of communication round $r$, the MBS updates the global model by
\begin{equation}\label{eq:global}
{g_r} = \frac{{\sum\nolimits_{b \in {\cal B}} {{K_b}\theta _t^b} }}{K}, t=\tau,
\end{equation}
where $K = \sum\nolimits_{b \in {\cal B}} {{K_b}}$ is the total amount of training data. After updating the global model $g_r$, the MBS will broadcast the global mode $g_{r}$ to all mSBSs to update their local models.

Hence, the workflow for the proposed BMFL is described as follows. Each mSBS (local agent) conducts local training for a deep neural network to predict Q value, which is used for guiding action decisions (which sectors should be covered) in the reinforcement framework. After each round of local training, the weights in the neural network from all local agents should be aggregated by the MBS based on (16) to update the global model, which then is shared with all the local agents to guide them in obtaining a more accurate deep neural network as per the rule in (10). The BMFL algorithm for beam management is presented as {\bf{Algorithm 1}}. As each communication round includes the computation of mSBS data cleaning and local model updating, the computational complexity of the proposed algorithm is given by ${\cal O}_{\rm{BMFL}} = {\cal O}({J \cdot ( {B \cdot U + \tau } )})$, where $J$ denotes the number of communication rounds. In fact, for BMFL, the local update step can be regarded as fully distributed DRL, as individual mSBSs train a local learning model based on local dataset without data interaction or aggregation (i.e., each mSBS performs gradient descent to adjust the local model parameter to minimize the loss function defined on its own dataset). The global aggregation step can be regarded as centralized DRL if the global aggregation is performed after every local update and the data samples and features are available for the aggregator (i.e., the MBS). However, the BMFL algorithm has greater advantages in privacy protection than traditional centralized and distributed DRL algorithms, due to the data cleaning and non-raw-data aggregation.

\begin{algorithm}[t]
  \caption{BMFL Algorithm}
  \hspace*{0.02in}{\bf Input:}
  ${\cal U}$, ${\cal B}$, ${\cal U}_b^*=\varnothing$, $s^{b}$, $a^{b}$, $\eta$, $\rho_b$, $\gamma$, $C$, $K_b$, $\lambda$, $\tau$\\
   \hspace*{0.02in}{\bf output:}
    Beam sectors $\mathcal{\pi}_{b}$.
  \begin{algorithmic}[1]
    \STATE Initialize experience relay pool $D_b, \forall b \in {\cal B}$;
    \STATE Initialize the global weights $g_0$;
    \FOR {communication round $r=1,2,...,J$}
    \STATE \underline{$\triangleright$Data cleaning}
    \FOR {$b=1,2,...,\left| {\cal B} \right|$}
    \FOR {$u=1,2,...,\left| {\cal U} \right|$}
    \IF {$d_{u,b}\leq \rho_b$ and {$\frac{n_u}{N_{total}}\leq \eta$}}
    \STATE ${\cal U}_b^*=\{{\cal U}_b^*, u\}$;
    \STATE $u=u+1$;
    \ELSE
    \STATE ${\cal U}_b^*=\{{\cal U}_b^*\}$;
    \STATE $u=u+1$;
    \ENDIF
    \ENDFOR
    \ENDFOR
    \STATE Collect data from  ${\cal U}_b^*$ for mSBS $b$, $\forall b\in {\cal B}$;
    \STATE \underline{$\triangleright$ Update local model}
      \IF {$r==1$}
    \STATE Initialize $\theta_0^{b}$, $\forall b\in {\cal B}$;
    \ELSE
    \STATE ~~~~$\theta_{0}^{b}=g_{r-1}-\frac{\lambda}{K_b}\sum_{b=1}^{\left| {{\cal B}} \right|}\nabla L(\theta_{t}^{b}), \forall b\in {\cal B}$.
    \ENDIF
    \STATE \underline{$\triangleright$ Train local model}
    \STATE Let $\hat{\theta}_0^{b}=\theta_0^{b}$, initialize target action-value function $\hat{Q}(\cdot)$ according to $\hat{\theta}_0^{b}$;
    \FOR {$t=1$ to $\tau$}
    \STATE Receive the initial state $s_{t}^{1},s_{t}^{2},...,s_{t}^{\left| {{\cal B}} \right|}$;
    \STATE Select $a_{t}^{b}={\text{argmax}_{a}Q(\cdot)}$ using $\epsilon$-greedy policy;
    \STATE Execute action $a_{t}^{b}$;
    \STATE Obtain ${\cal R}_{t}^{b}$ and $s_{t+1}^{b}$;
    \STATE Store $(s_{t}^{b},a_{t}^{b}, {\cal R}_{t}^{b},s_{t+1}^{b})$ into $D_b,\forall b \in {\cal B}$;
    \STATE Randomly select a sample $\left(s_{j}^{b},a_{j}^{b},{\cal R}_{j}^{b},s_{j+1}^{b}\right)$ from $D_b, 1 \le j \le t, \forall b \in {\cal B}$;
    \STATE Calculate $y_{t}^{b}$ according to equation (11);
    \STATE Perform a gradient descent step on $(y_{t}^{b}-Q(s_{t}^{b},a_{t}^{b};\theta_t^{b}))^{2}$;
    \STATE Update the parameter $\theta_t^{b}$ according to equation (15);
    \STATE Reset $\hat{Q}=Q$ every $C$ steps;
    \ENDFOR
    \STATE \underline{$\triangleright$ Update global model}
    \STATE $g_{r}=\frac{\sum_{b=1}^{|{\cal B}|}K_b\theta_{\tau}^{b}}{K}.$
    \ENDFOR

  \STATE Obtain beam sectors $\mathcal{\pi}_{b}$.
  \end{algorithmic}
\end{algorithm}

\begin{figure}[t]
  \begin{center}
    \scalebox{0.8}[0.8]{\includegraphics{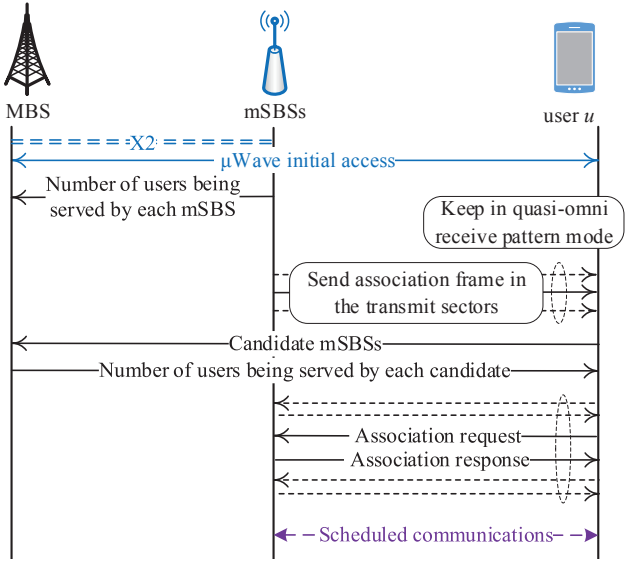}}
    \caption{Procedure of $\mu$Wave-assisted mmWave association.}
    \label{fig:5}
  \end{center}
\end{figure}

\subsection{User Association in UDmmN}
We propose a $\mu$Wave-assisted user association in UDmmN, where the information exchange is realized with the aid of $\mu$Wave as shown in Fig.~\ref{fig:5}. The details are as follows.

\emph{1) Downlink measurements:} In the association process, since mSBS $b$ ($\forall b \in {\cal B}$) has no information about idle users (or the users associated with other mSBSs) in either beam steering directions or the signal attenuation, the mSBS explores its antenna elements to form a sweeping beam to the users, while users operate in an omni receive pattern mode or quasi-omni mode (\emph{i.e.}, closely approximating the omni mode) to listen for the association frames broadcast by the mSBS. Note that mSBS $b$ only sweeps the transmit sectors in $\pi_b(t)$ at time $t$.

\emph{2) User perception:} After receiving the association frames, user $u$ ($u \in {\cal U}$) can know the mSBSs as well as the corresponding beam indexes. By measuring the received signal power, the user determines the set of the candidate serving mSBSs at time $t$, denoted by ${\cal B}_u^{\rm{can}} \left( t \right)$ (${\cal B}_u^{\rm{can}}\left( t \right) \subseteq {\cal B}$), and sends ${\cal B}_u^{\rm{can}} \left( t \right)$ to the MBS. The perceived signal power of user $u$ from beam $i$ of mSBS $b$ ($b \in {\cal B}_u^{\rm{can}} \left( t \right)$, $i \in \pi_{b,u}(t) \subseteq \pi_b(t)$) satisfies  ${\zeta _{u,b,i}}\left( t \right) \ge {\varsigma}$, where $\varsigma$ is a given threshold of the power, and $\pi_{b,u}(t)$ is the set of the candidate transmit sectors of mSBS $b$ for user $u$.

\emph{3) Multiple association:} For an idle user $u$, \emph{i.e.}, ${\cal B}_u (t-1) = \emptyset$, the association process can be outlined as follows.
\begin{itemize}
\item {\bf{repeat}}
\item Request to associate with mSBS $b^*$ ($b^* \in {\cal B}_u^{\rm{can}} (t)$) that meets
    ${\zeta _{u,{b^*},{i^*}}}(t) = \mathop {\max }\limits_{\forall b \in {\cal B}_u^{\rm{can}}(t),\forall i \in \pi_{b,u}(t)} {\zeta _{u,b,i}}(t)$;
\item Record mSBS $b^*$ into ${\cal B}_u(t)$ if ${x_{u,b}(t)} = 1$;
\item Remove mSBS $b^*$ out of ${\cal B}_u^{\rm{can}}(t)$;
\item {\bf{until}} $B_u(t) = \left| {{{\cal B}_u}(t)} \right| = B_u^{\max}$ or ${\cal B}_u^{\rm{can}} (t) = \emptyset$
\end{itemize}

For user $u$ of which ${\cal B}_u(t-1) \ne \emptyset$, some handover steps may be required before the above association. Moreover, we may jointly consider the perceived signal power ${\zeta _{u,b,i}}\left( t \right)$ and user load balancing between mSBSs when dealing with user association. For instance, user $u$ can choose the optimal candidate serving mSBS by calculating the variable ${y_{u,b,i}}\left( t \right)$ which is defined as ${y_{u,b,i}}\left( t \right) = {k_1}\frac{{{\zeta _{u,b,i}}\left( t \right)}}{\varsigma} + {k_2}\frac{U}{{\sum\nolimits_{u \in {\cal U}} {{x_{u,b}}\left( t \right)} }}$, where $k_1 $ and $k_2$ are the proportion factors, $0 < {k_1},{k_2} < 1$, and $k_1 + k_2 = 1$. The higher the value of $y_{u,b,i}$, the higher the priority of mSBS $b$ selected by user $u$. This method may overcome the problem of unbalanced mSBS loads which in turn affect the network fairness and may result in overly frequent handovers between
the adjacent mSBSs. It is left as our future work.

Meanwhile, as a coarse-grained beam training between user $u$ ($u \in {\cal U}$) and mSBS $b$ ($b \in {{{\cal B}_u}\left( t \right)}$) has been carried out during the above step 1) and 2), user $u$ can be served by beam $i^*$ of mSBS $b^*$ for the follow-up mmWave communication. For users with high QoS requirements, they may use directional beams to receive signals from the serving mSBSs, and thus a fine-grained beam training will be further required before data transmission. Since beam training is not the focus of this paper, we will not explain it in detail here.

\begin{table}[htbp]
\centering
\caption{Key simulation parameters.}
\scalebox{0.88}{
\begin{tabular}{c|c}
\Xhline{0.5pt}
\Xhline{0.5pt}
\textbf{Parameters} & \textbf{Values}\\
\Xhline{0.5pt}
\Xhline{0.5pt}
Carrier frequency of the MBS & $f_{c1} = 2.1$ GHz\\
Carrier frequency of mSBSs & $f_{c2} = 28$ GHz\\
Bandwidth of the MBS & $W_{mbs} = 100$ MHz\\
mmWave bandwidth & $W_{mm} = 2$ GHz\\
Number of transmit sectors of mSBS $b$ & $S_b =8$\\
Number of beams of mSBS $b$ & $M_b = 3$\\
Maximum number of mSBSs serving user $u$ & $B_u^{\max} = 3$\\
Transmit power of the MBS &  $p_u^0 = 50$ dBm\\
Transmit power of mSBSs & $p_{u,b} = 37$ dBm\\
Parameters of microwave channel gain & $\kappa =38.8$, $\rho = 2$\\
Parameters of mmWave path loss & $\alpha = 61.3$, $\beta = 2.1$,\\
  & $\xi \sim N\left( {0,4} \right)$\\
mmWave transmit antenna gain  & $G _{u,b}^T = 12$ dB\\
mmWave receive antenna gain & $G_{u,b}^R = 10$ dB\\
\Xhline{0.5pt}
Number of layers in neural network & 4\\
Target network update interval step & 4\\
Discount factor  &  0.8\\
Learning rate for training & 0.1\\
Replay memory size & 400\\
Minibatch size & 36\\
\Xhline{0.5pt}
\Xhline{0.5pt}
\end{tabular}
}
\end{table}

\begin{table}[htbp]\small
\centering
\caption{Structure of the neural network.}
\scalebox{0.9}{
\begin{tabular}{c|c}
\Xhline{0.5pt}
\Xhline{0.5pt}
\textbf{Layer} & \textbf{Generate}\\
\Xhline{0.5pt}
\Xhline{0.5pt}
1	& nn.Sequential(nn.Linear(in\_dim, n\_hidden\_1), nn.ReLU())\\
2	& nn.Sequential(nn.Linear(n\_hidden\_1, n\_hidden\_2), nn.ReLU())\\
3	& nn.Sequential(nn.Linear(n\_hidden\_2, n\_hidden\_3), nn.ReLU())\\
4	& nn.Sequential(nn.Linear(n\_hidden\_3, out\_dim))\\
\Xhline{0.5pt}
\Xhline{0.5pt}
\end{tabular}
}
\end{table}

\section{Performance Evaluation}
In our simulation, we consider a square area with the size $100m \times 100m$, where an MBS is located at the center, multiple mSBSs are distributed within the macro cell uniformly, and multiple users are randomly distributed in the cell. To make the results convincing, we compare the performance of the proposed algorithm with other schemes under the same user distribution. For the coverage of each mSBS, we uniformly divide it into 8 sectors (\emph{i.e.}, each sector is with the coverage of ${45}^{\circ}$). Each mSBS generates three beams covering different sectors. Each user can be associated with up to 3 mSBSs. TABLE II summarizes the detailed simulation parameters. Assuming that the MBS bandwidth is evenly allocated to the serving users, i.e., the macro wave bandwith for user $u$ is given by $W_{u, mbs} = \frac{W_{mbs}} {N_{mbs}}$, it can be considered that there is no cochannel interference between users. Meanwhile, the mmWave bandwith for user $u$ is assumed to be $W_{u,b} = W_{mm}$. As both the mSBSs and users in mmWave network usually transmit/receive signals with directional beams pointing in different angular directions (spatially orthogonal to each other), the interference between simultaneous mmWave links will only be caused by beam sidelobes, which is usually very small and negligible. Furthermore, the co-beam interference (the interference between the users served by the same mSBS transmit beam) may be eliminated by appropriate spatial precoding. Therefore, the SINR can be approximated by the SNR for mmWaves which is quite different than that in sub-6GHz networks. In the learning part settings, we consider a fully connected neural network for each mSBS. We mainly use the nn.Module, nn.Sequential, and nn.Linear of PyTorch to build a four-layer neural network of which the structure is given in TABLE III, where in\_dim =$\left| {\cal U} \right| + 2 M_b + 1$, n\_hidden\_1 = 40, n\_hidden\_2 = 60, n\_hidden\_3 = 40, and out\_dim = 1. Specifically, the two parameters in nn.Linear() are the size of each input sample and the size of each output sample respectively. Meanwhile, the size of experience replay pool for each mSBS is set to 400, the batch-size is set to 36, and the learning rate is set to 0.1. All the other settings such as reward, action, state keep consistent with those in our modeling part.

\begin{figure}[t]
  \begin{center}
    \scalebox{0.6}[0.6]{\includegraphics{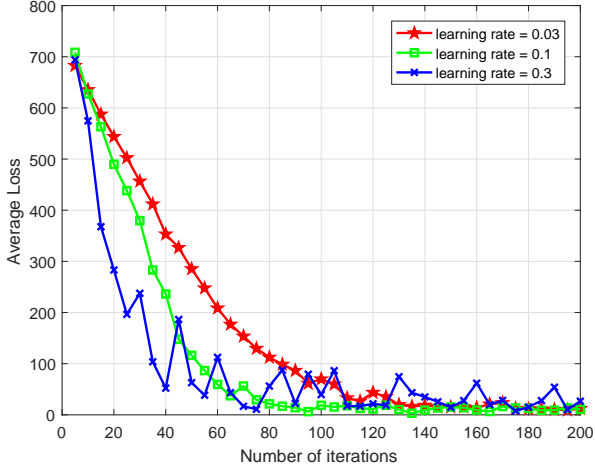}}
    \caption{Convergence of the BMFL in UDmmN.}
    \label{fig:6}
  \end{center}
\end{figure}

We first evaluate the convergence performance of the proposed BMFL algorithm in terms of the average loss function value, as shown in Fig.~\ref{fig:6}. From this figure, we find that all curves under the three typical learning rates, 0.03, 0.1 and 0.3, reach the convergence after a certain number of iterations. Specifically, the BMFL algorithm reaches the convergence after around 80 iterations when learning rate is 0.1 while around 130 iterations of learning rate 0.03 and nearly 200 iterations for learning rate 0.3. These convergence results clearly demonstrate the effectiveness and rationality of BMFL. Meanwhile, observed from the figure, the average loss value of the algorithm presents dramatical decreasing in the beginning stages. The reason lies in the fact that, the gradient descent approach is employed by the BMFL algorithm to train the DDQN-based optimization framework, thereby the loss function converges faster at the beginning while becoming gentle near the minimum point. In addition to the learning rate, the convergence performance of BMFL is also related to some other hyperparameters such as the size of replay memory and minibatches. The parameters in TABLE II are chosen to make a tradeoff between the communication performance and the computational complexity according to the simulation results that are not shown here. These settings may not be optimal, but they can make our BMFL algorithm achieve a good convergence performance which will stimulate its practical application.

We then conduct numerical simulations to compare the performance in terms of user coverage and network throughput with the following four beam management schemes.

\hangafter 1
\hangindent 1.0em
\noindent
1) Brute-Force Search (BFS): Find the optimal beam coverage by searching all the possible beam sectors. This algorithm can reach the optimal solution of beam management with extremely high computational complexity.

\hangafter 1
\hangindent 1.0em
\noindent
2) Evenly Deployed Beam (EDB): Deploy the beams in a uniform manner. In EDB, we only need to optimize the direction of one beam for each mSBS, and the direction of the other beams can thus be determined as the rule of uniform deployment.

\hangafter 1
\hangindent 1.0em
\noindent
3) Beam Management based on Distributed Learning (BMDL): Individual mSBSs train their own data through DDQN and make decision on beam configuration independently, where no data aggregation of FL is used.

\hangafter 1
\hangindent 1.0em
\noindent
4) Beam Management based on Centralized Learning (BMCL): All mSBSs transmit data to a controller (i.e., the MBS) for centralized training in DDQN. Then the MBS makes global decision on beam configuration for all mSBSs.

\begin{figure}[t]
  \begin{center}
    \scalebox{0.6}[0.6]{\includegraphics{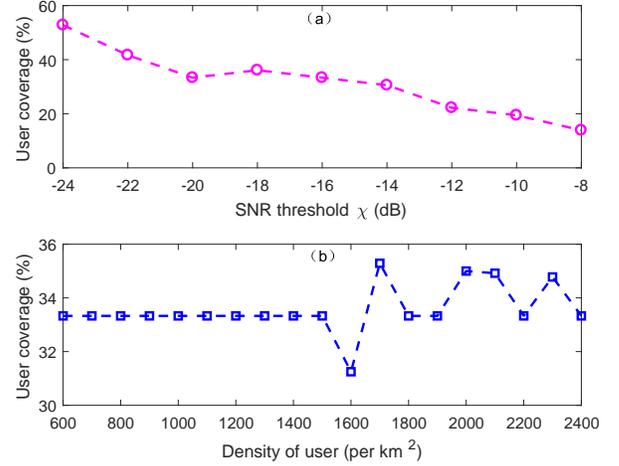}}
    \caption{User coverage performance for BMFL versus (a) SNR threshold, given that $B=\left| {\cal B} \right| = 3$ and $U=\left| {\cal U} \right| = 12$, and (b) user density, given that $B=\left| {\cal B} \right| = 6$ and $\chi$ = -20 dB.}
    \label{fig:7}
  \end{center}
\end{figure}

\begin{figure}[t]
  \begin{center}
    \scalebox{0.6}[0.6]{\includegraphics{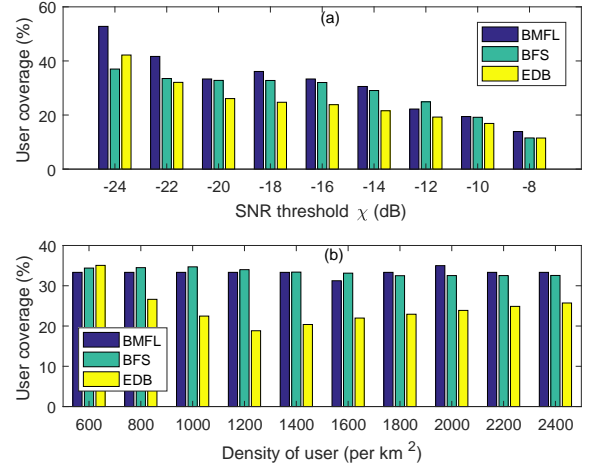}}
    \caption{Comparisons of user coverage for the BMFL, BFS and EDB, versus (a) SNR threshold, given that $B=\left| {\cal B} \right| = 3$ and $U=\left| {\cal U} \right| = 12$, and (b) user density, given that $B=\left| {\cal B} \right| = 6$ and $\chi$ = -20 dB.}
    \label{fig:8}
  \end{center}
\end{figure}

\subsection{Comparison with BFS and EDB}
In this subsection, we compare the performance of BMFL with that of the two traditional schemes, BFS and EDB.

\emph{User coverage}: We evaluate the user coverage performance for BMFL versus SNR threshold $\chi$ and user density $Den_{\rm{user}}$ respectively, as shown in Fig.~\ref{fig:7}. As multiple association is considered for users, the user coverage in our experiments is defined as $Cov_{\rm{user}} = {{\left( {\sum\nolimits_{\forall b \in {\cal B}} {{N_b}} } \right)} \mathord{\left/  {\vphantom {{\left( {\sum\nolimits_{\forall b \in {\cal B}} {{N_b}} } \right)} {\left[ {U \cdot \min \left( {B,B_u^{\max }} \right)} \right]}}} \right.  \kern-\nulldelimiterspace} {\left[ {U \cdot \min \left( {B,B_u^{\max }} \right)} \right]}}$. As a simple example, when $U=1$, $B=4$, and $B_u^{\max } = 3$, we have $Cov_{\rm{user}} = \frac{{1 + 0 + 1 + 0}}{{1 \times 3}} = 66.67 \%$ if the user is associated with mSBS 1 and mSBS 3 simultaneously. It can be seen from Fig.~\ref{fig:7}(a) that $Cov_{\rm{user}}$ decreases with $\chi$. For example, under the same condition, we get $Cov_{\rm{user}} = 52.8 \%$ when $\chi = -24$ dB, but $Cov_{\rm{user}} = 13.9 \%$ if $\chi = -8$ dB. A lower value of $\chi$ indicates that users may be served even in a poor signal environment, so as to obtain higher user coverage. By contrast, fewer users can be served if the value of $\chi$ is set high. Given that $\chi = -20$ dB, the result of $Cov_{\rm{user}}$ versus the user density $Den_{\rm{user}}$ is shown in Fig.~\ref{fig:7}(b). Here the user density is defined as $Den_{\rm{user}} = {U \mathord{\left/  {\vphantom {U {Area}}} \right. \kern-\nulldelimiterspace} {Area}}$, where $U=\left| {\cal U} \right|$ is the number of users in a certain area $Area$ at time $t$. For example, we have $Den_{\rm{user}} = {{20} \mathord{\left/  {\vphantom {{20} {\left( {0.1 \times 0.1} \right)}}} \right.  \kern-\nulldelimiterspace} {\left( {0.1 \times 0.1} \right)}} = 2000$ per $km^2$ when we set $U = 20$ in the simulation. It is obviously to see that, when the value of $Den_{\rm{user}}$ changes, the fluctuation of $Cov_{\rm{user}}$ is not large and is roughly stable between 31$\%$ and 35$\%$. We compare BMFL with the two traditional schemes BFS and EDB, as shown in Fig.~\ref{fig:8}. By analyzing these results, we can see that the performance of the proposed scheme in user coverage is generally consistent with the optimal scheme BFS, but better than EDB. Also worth noting is that for all the three schemes, the value of $Den_{\rm{user}}$ in the results is not very high. This is mainly because we divide the service range of mSBS $b$ ($\forall b \in {\cal B}$) into 8 sectors and only three operating beams, i.e., $S_b = 8$ and $M_b = 3$.

\begin{figure}[t]
  \begin{center}
    \scalebox{0.6}[0.6]{\includegraphics{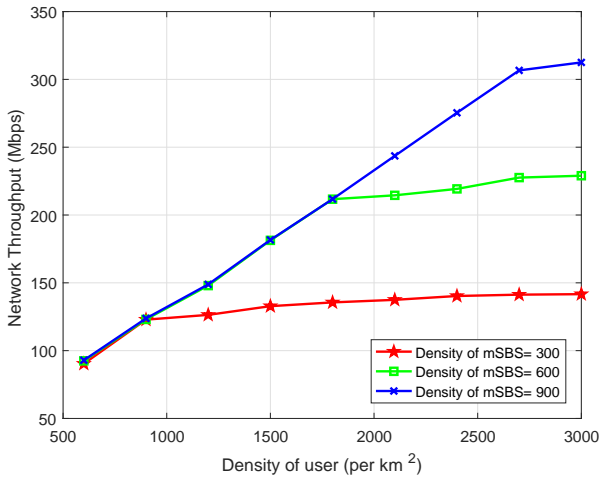}}
    \caption{Network throughput performance for BMFL under different mSBS densities, given that $\chi$ = -20 dB.}
    \label{fig:9}
  \end{center}
\end{figure}

\begin{figure}[t]
  \begin{center}
    \scalebox{0.6}[0.6]{\includegraphics{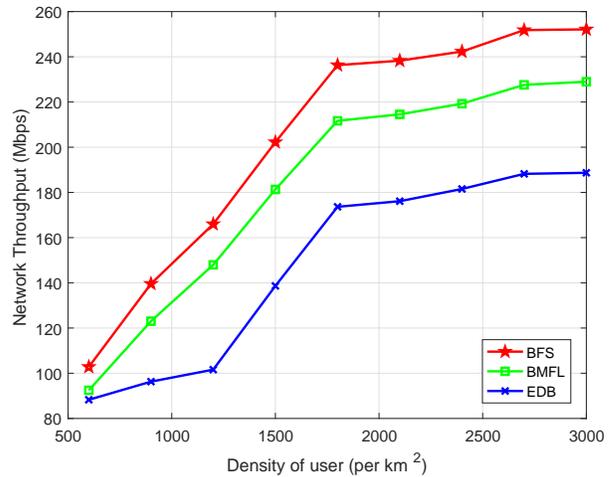}}
    \caption{Comparisons of network throughput for the BMFL, BFS and EDB, given that $B=\left| {\cal B} \right| = 6$ and $\chi$ = -20 dB.}
    \label{fig:10}
  \end{center}
\end{figure}

\emph{Network throughput}: We evaluate network throughput performance of BMFL with the varying density of user under three different mSBS densities, as shown in Fig.~\ref{fig:9}. Similar to the definition of user coverage, let $B=\left| {\cal B} \right|$ be the number of users in a certain area $Area$ at time $t$, then the mSBS density of this area can be given as $Den_{\rm{mSBS}} = {B \mathord{\left/  {\vphantom {B {Area}}} \right.  \kern-\nulldelimiterspace} {Area}}$. As expected, we find that the achieved network throughput $R$ usually increases with $Den_{\rm{user}}$ in each mSBS density. Meanwhile, the value of $R$ has little difference for all the three mSBS densities when the user density is low, e.g., when $Den_{\rm{user}} \leq 900$ per $km^2$, because the number of users served may not vary much at this time. With the increase of $Den_{\rm{user}}$ the value of $R$ is the largest under the highest mSBS density (i.e., when  $Den_{\rm{mSBS}} = 900$ per $km^2$) which is because of the abundant beam resource. That is, generally, the denser the mSBS and the more operating beams, the more users can be served, resulting in the greater network throughput.

Furthermore, we compare the network throughput of BMFL with BFS and EDB. Fig.~\ref{fig:10} shows the network throughput of the three beam management algorithms under the varying user density from 600 per $km^2$ to 3000 per $km^2$. We fix the mSBS density as 600 per $km^2$ in this simulation. As expected that the BFS beam management algorithm achieves the highest throughput as all the potential solutions have been searched and tested. Importantly, we find that BMFL achieves the second highest network throughput with relatively small difference of that in BFS but much higher than that of EDB. For example, when $Den_{\rm{user}} =3000$ per $km^2$, the network throughput of the three schemes are $R_{\rm{BFS}}  \approx 252 Mbps$, $R_{\rm{BMFL}}  \approx 229 Mbps$, and $R_{\rm{EDB}}  \approx 188 Mbps$, respectively. These results further demonstrate the performance gain of the proposed BMFL algorithm. Meanwhile, although the performance of the proposed scheme in terms of user coverage is comparable to that of the optimal scheme BFS, it is slightly weak in terms of network throughput, which may be due to the difference of users they serve. We believe that when there is enough training data, the performance of BMFL in terms of network throughput can also reach the level of the BFS.

\subsection{Comparison with BMDL and BMCL}
In this subsection, we compare the performance of BMFL with that of the two adaptive schemes, BMDL and BMCL. For BMDL and BMCL, the setting of simulation parameters (neural network, user distribution, etc.) is consistent with the proposed BMFL.

\begin{figure}[t]
  \begin{center}
    \scalebox{0.6}[0.6]{\includegraphics{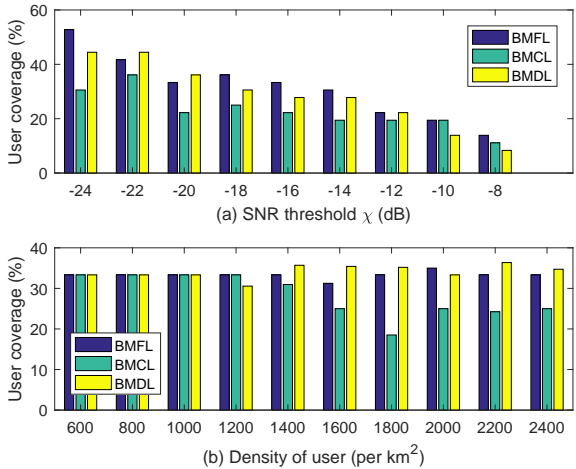}}
    \caption{Comparisons of user coverage for the BMFL, BMCL and BMDL, versus (a) SNR threshold, given that $B=\left| {\cal B} \right| = 3$ and $U=\left| {\cal U} \right| = 12$, and (b) user density, given that $B=\left| {\cal B} \right| = 6$ and $\chi$ = -20 dB.}
    \label{fig:11}
  \end{center}
\end{figure}

\begin{figure}[t]
  \begin{center}
    \scalebox{0.6}[0.6]{\includegraphics{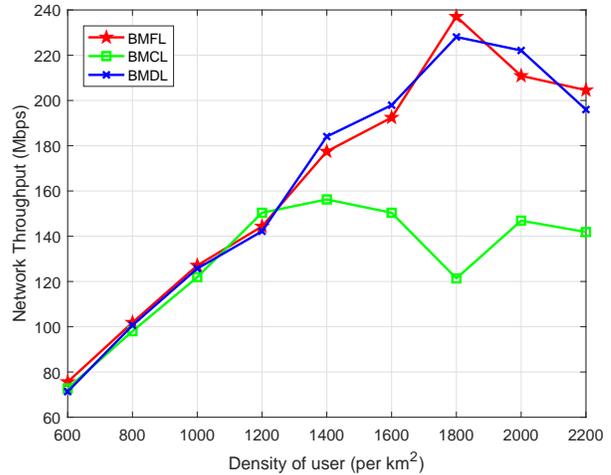}}
    \caption{Comparisons of network throughput for the BMFL, BMCL and BMDL, given that $B=\left| {\cal B} \right| = 6$ and $\chi$ = -20 dB.}
    \label{fig:12}
  \end{center}
\end{figure}

Fig.~\ref{fig:11} shows the performance comparison in terms of user coverage. When the SINR threshold $\chi$ is fixed, we see that the proposed BMFL can achieve a higher user coverage than BMDL and BMCL. For relatively low user density, e.g., when $Den_{\rm{user}} <1200$ per $km^2$, the user coverage of these three schemes is almost the same. With the increase of $Den_{\rm{user}}$, the user coverage performance of BMFL is not much different from that of BMDL, but better than that of BMCL. The variation of network throughput with user density is similar to that of the user coverage, as shown in Fig.~\ref{fig:12}. For example, when $Den_{\rm{user}} = 800$ per $km^2$, the network throughput of the three schemes has little difference, where $R_{\rm{BMFL}}  \approx R_{\rm{BMCL}} \approx R_{\rm{BMDL}}  \approx 100 Mbps$. For a high $Den_{\rm{user}}$, the throughput of BMFL and BMDL is usually close, but higher than that of BMCL. Compared with BMDL, the proposed BMFL also adopts a distributed learning architecture. However, it has an advantage that the training model accuracy and learning convergence speed can be improved through the cooperation of multi-agent (e.g., mSBSs). Due to the small number of mSBSs in the simulation, this advantage is not obvious, so the performance of the two is similar. For BMCL, it may be that the algorithm converges to a suboptimal solution, which leads to the low results. The essence of machine learning algorithms determines that the optimal result cannot be guaranteed. It is an iterative updating process.


\section{Conclusions}
Due to the directional transmission and dense network deployment, complexity of beam management problem in mmWave communication systems becomes a real challenge. To address the complex and dynamic beam control issue, in this paper we have proposed a federated DRL-based adaptive beam management algorithm, BMFL. In BMFL, individual mSBSs train a local machine learning model based on the cleaned local dataset and then send the model features to the MBS for aggregation. Meanwhile, we employed DDQN to train the local model on mSBSs under an FL framework. Due to the data cleaning and non-raw-data aggregation, the proposed BMFL algorithm has great advantages in privacy protection and wireless resource conservation (e.g., transmit power, bandwidth). Simulation results have shown that the BMFL provides a better tradeoff between computational complexity and network throughput. Moreover, the performance of the proposed scheme in user coverage is generally comparable to that of the optimal scheme BFS, which is also verified by the simulations. In general, this work can be seen as a pioneer of using FL to solve the systematic beam management problem under UDN scenarios.

\section*{Acknowledgment}
The work of Q. Xue was supported in part by NSFC under Grant 62001071, the Macao Young Scholars Program under Grant AM2021018, the China Postdoctoral Science Foundation under Grant 2020M683291, and the Science and Technology Research Program of Chongqing Municipal Education Commission under Grant KJQN202200617. The work of L. Yan was supported in part by NSFC under Grant 62101460, Project funded by China Postdoctoral Science Foundation under Grant 2019TQ0270, and Sichuan Provincial NSFC under Grant 2022NSFSC0893. The work of S. Ma was supported in part by the Science and Technology Development Fund, Macau SAR (File no. 0036/2019/A1 and File no. SKL-IOTSC(UM)-2021-2023); in part by the Research Committee of University of Macau under Grant MYRG2020-00095-FST.

\ifCLASSOPTIONcaptionsoff
  \newpage
\fi

\bibliographystyle{IEEEtran}
\bibliography{reference}

\begin{thebibliography}{10}
\providecommand{\url}[1]{#1}
\csname url@samestyle\endcsname
\providecommand{\newblock}{\relax}
\providecommand{\bibinfo}[2]{#2}
\providecommand{\BIBentrySTDinterwordspacing}{\spaceskip=0pt\relax}
\providecommand{\BIBentryALTinterwordstretchfactor}{4}
\providecommand{\BIBentryALTinterwordspacing}{\spaceskip=\fontdimen2\font plus
\BIBentryALTinterwordstretchfactor\fontdimen3\font minus
  \fontdimen4\font\relax}
\providecommand{\BIBforeignlanguage}[2]{{%
\expandafter\ifx\csname l@#1\endcsname\relax
\typeout{** WARNING: IEEEtran.bst: No hyphenation pattern has been}%
\typeout{** loaded for the language `#1'. Using the pattern for}%
\typeout{** the default language instead.}%
\else
\language=\csname l@#1\endcsname
\fi
#2}}
\providecommand{\BIBdecl}{\relax}
\BIBdecl

\bibitem{Globecom-BMFL}
J.~{Wang}, Q.~{Xue}, Y.~{Sun}, G.~{Feng}, L.~{Tang}, and S.~{Ma}, ``Beam
  management in ultra-dense millimeter wave network via federated learning,''
  in \emph{2021 IEEE Global Communications Conference: Mobile and Wireless
  Networks}, 2021, pp. 1--6.

\bibitem{Cisco}
Cisco, ``Cisco annual internet report (2018-2023),'' \emph{White paper}, March,
  2020.

\bibitem{Two-Way-Massive-MIMO-Relaying}
J.~Feng, S.~Ma, S.~A{\"i}ssa, and M.~Xia, ``Two-way massive {MIMO} relaying
  systems with non-ideal transceivers: Joint power and hardware scaling,''
  \emph{IEEE Transactions on Communications}, vol.~67, no.~12, pp. 8273--8289,
  2019.

\bibitem{Tensor-mmWave-MIMO}
X.~Wu, S.~Ma, and X.~Yang, ``Tensor-based low-complexity channel estimation for
  mm{W}ave massive {MIMO-OTFS} systems,'' \emph{Journal of Communications and
  Information Networks}, vol.~5, no.~3, pp. 324--334, 2020.

\bibitem{Matrix-Monotonic-Optimization}
C.~Xing, S.~Wang, S.~Chen, S.~Ma, H.~V. Poor, and L.~Hanzo, ``Matrix-monotonic
  optimization $-$ part {I}: Single-variable optimization,'' \emph{IEEE
  Transactions on Signal Processing}, vol.~69, pp. 738--754, 2021.

\bibitem{ultra-dense-networks}
M.~{Kamel}, W.~{Hamouda}, and A.~{Youssef}, ``Ultra-dense networks: A survey,''
  \emph{IEEE Communications Surveys $\&$ Tutorials}, vol.~18, no.~4, pp.
  2522--2545, 2016.

\bibitem{Multi-Connectivity}
A.~{Wolf}, P.~{Schulz}, M.~{D\"orpinghaus}, J.~C.~S. {Santos Filho}, and
  G.~{Fettweis}, ``How reliable and capable is multi-connectivity?'' \emph{IEEE
  Transactions on Communications}, vol.~67, no.~2, pp. 1506--1520, 2019.

\bibitem{Large-Intelligent-Surface-Antennas-LISA}
Y.-C. Liang, R.~Long, Q.~Zhang, J.~Chen, H.~V. Cheng, and H.~Guo, ``Large
  intelligent surface/antennas ({LISA}): Making reflective radios smart,''
  \emph{Journal of Communications and Information Networks}, vol.~4, no.~2, pp.
  40--50, 2019.

\bibitem{Unified-IRS-Aided-MIMO}
S.~Gong, C.~Xing, X.~Zhao, S.~Ma, and J.~An, ``Unified {IRS}-aided {MIMO}
  transceiver designs via majorization theory,'' \emph{IEEE Transactions on
  Signal Processing}, vol.~69, pp. 3016--3032, 2021.

\bibitem{Rate-Analysis-RIS-Assisted-MIMO}
J.~Zhang, J.~Liu, S.~Ma, C.-K. Wen, and S.~Jin, ``Large system achievable rate
  analysis of {RIS}-assisted {MIMO} wireless communication with statistical
  {CSIT},'' \emph{IEEE Transactions on Wireless Communications}, vol.~20,
  no.~9, pp. 5572--5585, 2021.

\bibitem{Sum-Rate-RIS-Assisted-MIMO}
K.~Xu, J.~Zhang, X.~Yang, S.~Ma, and G.~Yang, ``On the sum-rate of
  {RIS}-assisted {MIMO} multiple-access channels over spatially correlated
  rician fading,'' \emph{IEEE Transactions on Communications}, vol.~69, no.~12,
  pp. 8228--8241, 2021.

\bibitem{Tutorial-Beam-Management}
M.~{Giordani}, M.~{Polese}, A.~{Roy}, D.~{Castor}, and M.~{Zorzi}, ``A tutorial
  on beam management for 3{GPP} {NR} at mm{W}ave frequencies,'' \emph{IEEE
  Communications Surveys \& Tutorials}, vol.~21, no.~1, pp. 173--196, 2019.

\bibitem{Federated-Learning}
S.~{Niknam}, H.~S. {Dhillon}, and J.~H. {Reed}, ``Federated learning for
  wireless communications: Motivation, opportunities, and challenges,''
  \emph{IEEE Communications Magazine}, vol.~58, no.~6, pp. 46--51, 2020.

\bibitem{Modular-High-Resolution}
E.~{Onggosanusi}, M.~S. {Rahman}, L.~{Guo}, Y.~{Kwak}, H.~{Noh}, Y.~{Kim},
  S.~{Faxer}, M.~{Harrison}, M.~{Frenne}, S.~{Grant}, R.~{Chen}, R.~{Tamrakar},
  and a.~Q.~{Gao}, ``Modular and high-resolution channel state information and
  beam management for 5{G} new radio,'' \emph{IEEE Communications Magazine},
  vol.~56, no.~3, pp. 48--55, 2018.

\bibitem{Standalone-Non-Standalone}
M.~{Giordani}, M.~{Polese}, A.~{Roy}, D.~{Castor}, and M.~{Zorzi}, ``Standalone
  and non-standalone beam management for 3{GPP} {NR} at mm{W}aves,'' \emph{IEEE
  Communications Magazine}, vol.~57, no.~4, pp. 123--129, 2019.

\bibitem{Beam-Management-5G-Beyond}
Y.~R. {Li}, B.~{Gao}, X.~{Zhang}, and K.~{Huang}, ``Beam management in
  millimeter-wave communications for 5{G} and beyond,'' \emph{IEEE Access},
  vol.~8, pp. 13\,282--13\,293, 2020.

\bibitem{Beam-Management-Q-Learning}
D.~C. {Ara¨²jo} and A.~L.~F. {de Almeida}, ``Beam management solution using
  {Q}-learning framework,'' in \emph{2019 IEEE 8th International Workshop on
  Computational Advances in Multi-Sensor Adaptive Processing (CAMSAP)}, 2019,
  pp. 594--598.

\bibitem{802.11ay}
Y.~{Ghasempour}, C.~R. C.~M. {da Silva}, C.~{Cordeiro}, and E.~W. {Knightly},
  ``I{EEE} 802.11ay: Next-generation 60 {GH}z communication for 100 {G}b/s
  {W}i-{F}i,'' \emph{IEEE Communications Magazine}, vol.~55, no.~12, pp.
  186--192, 2017.

\bibitem{Beamspace-SU-MIMO}
Q.~Xue, X.~Fang, and C.-X. Wang, ``Beamspace {SU-MIMO} for future millimeter
  wave wireless communications,'' \emph{IEEE Journal on Selected Areas in
  Communications}, vol.~35, no.~7, pp. 1564--1575, 2017.

\bibitem{Beam-Management-Beamspace-MU-MIMO}
Q.~{Xue}, X.~{Fang}, M.~{Xiao}, S.~{Mumtaz}, and J.~{Rodriguez}, ``Beam
  management for millimeter-wave beamspace {MU}-{MIMO} systems,'' \emph{IEEE
  Transactions on Communications}, vol.~67, no.~1, pp. 205--217, 2019.

\bibitem{Scalable-Interference-Coordination}
W.~Feng, Y.~Wang, D.~Lin, N.~Ge, J.~Lu, and S.~Li, ``When mm{W}ave
  communications meet network densification: A scalable interference
  coordination perspective,'' \emph{IEEE Journal on Selected Areas in
  Communications}, vol.~35, no.~7, pp. 1459--1471, 2017.

\bibitem{Tracking-angles-departure-arrival}
C.~{Zhang}, D.~{Guo}, and P.~{Fan}, ``Tracking angles of departure and arrival
  in a mobile millimeter wave channel,'' in \emph{2016 IEEE International
  Conference on Communications (ICC)}, 2016, pp. 1--6.

\bibitem{Robust-Beam-Tracking}
S.~{Jayaprakasam}, X.~{Ma}, J.~W. {Choi}, and S.~{Kim}, ``Robust beam-tracking
  for mmwave mobile communications,'' \emph{IEEE Communications Letters},
  vol.~21, no.~12, pp. 2654--2657, 2017.

\bibitem{Codebook-Based-Training}
D.~{Zhang}, A.~{Li}, M.~{Shirvanimoghaddam}, P.~{Cheng}, Y.~{Li}, and
  B.~{Vucetic}, ``Codebook-based training beam sequence design for
  millimeter-wave tracking systems,'' \emph{IEEE Transactions on Wireless
  Communications}, vol.~18, no.~11, pp. 5333--5349, 2019.

\bibitem{beam-Bandit-Learning}
J.~{Zhang}, Y.~{Huang}, Y.~{Zhou}, and X.~{You}, ``Beam alignment and tracking
  for millimeter wave communications via bandit learning,'' \emph{IEEE
  Transactions on Communications}, vol.~68, no.~9, pp. 5519--5533, 2020.

\bibitem{sun2017smart}
Y.~Sun, G.~Feng, S.~Qin, Y.-C. Liang, and T.-S.~P. Yum, ``The {SMART} handoff
  policy for millimeter wave heterogeneous cellular networks,'' \emph{IEEE
  Transactions on Mobile Computing}, vol.~17, no.~6, pp. 1456--1468, 2018.

\bibitem{Deep-Learning-Coordinated-Beamforming}
A.~{Alkhateeb}, S.~{Alex}, P.~{Varkey}, Y.~{Li}, Q.~{Qu}, and D.~{Tujkovic},
  ``Deep learning coordinated beamforming for highly-mobile millimeter wave
  systems,'' \emph{IEEE Access}, vol.~6, pp. 37\,328--37\,348, 2018.

\bibitem{Deep-Learning--Beam-Management-Interference}
P.~{Zhou}, X.~{Fang}, X.~{Wang}, Y.~{Long}, R.~{He}, and X.~{Han}, ``Deep
  learning-based beam management and interference coordination in dense
  mm{W}ave networks,'' \emph{IEEE Transactions on Vehicular Technology},
  vol.~68, no.~1, pp. 592--603, 2019.

\bibitem{Deep-learning-channel-estimation-tracking}
S.~{Moon}, H.~{Kim}, and I.~{Hwang}, ``Deep learning-based channel estimation
  and tracking for millimeter-wave vehicular communications,'' \emph{Journal of
  Communications and Networks}, vol.~22, no.~3, pp. 177--184, 2020.

\bibitem{User-centric-Association}
Q.~Xue, Y.~Sun, J.~Wang, G.~Feng, L.~Yan, and S.~Ma, ``User-centric association
  in ultra-dense mm{W}ave networks via deep reinforcement learning,''
  \emph{IEEE Communications Letters}, vol.~25, no.~11, pp. 3594--3598, 2021.

\bibitem{FL-Edge-Computing}
S.~Wang, T.~Tuor, T.~Salonidis, K.~K. Leung, C.~Makaya, T.~He, and K.~Chan,
  ``Adaptive federated learning in resource constrained edge computing
  systems,'' \emph{IEEE Journal on Selected Areas in Communications}, vol.~37,
  no.~6, pp. 1205--1221, 2019.

\bibitem{RAN-Slicing-FL}
Y.-J. Liu, G.~Feng, Y.~Sun, S.~Qin, and Y.-C. Liang, ``Device association for
  {RAN} slicing based on hybrid federated deep reinforcement learning,''
  \emph{IEEE Transactions on Vehicular Technology}, vol.~69, no.~12, pp.
  15\,731--15\,745, 2020.

\bibitem{DDQN}
H.~v. Hasselt, A.~Guez, and D.~Silver, ``Deep reinforcement learning with
  double {Q}-learning,'' in \emph{AAAI'16: Proceedings of the Thirtieth {AAAI}
  Conference on Artificial Intelligence}, 2016, pp. 2094--2100.

\bibitem{A-Learning-Approach}
C.~{Shen} and M.~{van der Schaar}, ``A learning approach to frequent handover
  mitigations in 3{GPP} mobility protocols,'' in \emph{2017 IEEE Wireless
  Communications and Networking Conference (WCNC)}, 2017, pp. 1--6.

\bibitem{Microwave-channel-model}
S.~{Zang}, W.~{Bao}, P.~L. {Yeoh}, B.~{Vucetic}, and Y.~{Li}, ``Managing
  vertical handovers in millimeter wave heterogeneous networks,'' \emph{IEEE
  Transactions on Communications}, vol.~67, no.~2, pp. 1629--1644, 2019.

\bibitem{Channel-Modeling}
M.~R. {Akdeniz}, Y.~{Liu}, M.~K. {Samimi}, S.~{Sun}, S.~{Rangan}, T.~S.
  {Rappaport}, and E.~{Erkip}, ``Millimeter wave channel modeling and cellular
  capacity evaluation,'' \emph{IEEE J. Sel. Areas Commun.}, vol.~32, no.~6, pp.
  1164--1179, 2014.

\end{thebibliography}

\end{document}